\documentclass[12pt,a4paper]{article}
\begin{document}

\def\br{{\bf r}}
\def\bQ{{\bf Q}}

\title{\bf Kerr constant of vesicle-like droplets}
\author{M. Richterov\'{a} and V. Lis\'{y}\\\\
Department of Biophysics, P. J. \v{S}af\'{a}rik University,\\
Jesenn\'{a} 5, 041 54 Ko\v{s}ice, Slovakia}
\date{}
\maketitle

\begin{abstract}
\noindent   The Kerr effect on vesicle-like droplets is described.
We give a simple derivation of the Kerr constant for a dielectric
fluid droplet immersed in another fluid, assuming that the droplet
in an electric field becomes a prolate ellipsoid. The Kerr
constant is evaluated also for a droplet covered by a membrane of
finite thickness. The latter result differs significantly from the
result by E. van der Linden \textit{et al.} [Physica A {\bf 156},
130 (1989)]. Due to the difference the bending rigidity constant
of the surface layer extracted from their experiments on droplet
microemulsions should be increased about two times.

\end{abstract}

\section*{Introduction}

The vesicle~\cite{mic,modern,seifert} is a closed surface of the
lipid-bilayer membrane. It is enclosed by a fluid and, in general,
is filled inside with another fluid. The membrane is several nm in
thickness and exists as a biological membrane. A cell, typically
$\sim10$ nm in diameter, itself can be regarded as a vesicle, and
usually contains intracellular vesicles as organelles. Artificial
vesicles, called liposomes, can carry drugs in their interior, and
are used in the experimental biology and pharmacology. Within the
phenomenological "biologically inspired" physical models the
formation (thermodynamics) and the dynamics of vesicles are
described neglecting the membrane thickness when compared to its
radius. In such an approach also the microemulsion droplets coated
with a surfactant monolayer film can be studied simultaneously. We
shall refer to all the mentioned fluid droplets as to vesicle-like
droplets. Their surface membranes or films can be in the frame of
the Canham-Helfrich model of interfacial
elasticity~\cite{canham,helfrich} characterized by a few
parameters: the bending and saddle splay modules $\kappa$ and
$\bar{\kappa}$, the spontaneous curvature radius $R_{s}$, the
surface tension coefficient $\sigma$, and the equilibrium radius
$R$. The determination of these constants has been attempted by
several macroscopic and microscopic methods. However, different
techniques yield very different parameters of the surface
shells~\cite{lisybru}. It is thus important to have adequate
theoretical description of the experimental techniques that could
serve as alternative probes of the vesicle membrane
properties.\\
   With this aim, in the present work the theory of the Kerr effect
on vesicle-like droplets is developed. We give a simple derivation
of the Kerr constant for a fluid droplet having a spherical shape
in equilibrium. First we assume that the membrane thickness is
negligible if compared to the vesicle radius. Placing such a
droplet in a constant electric field, the vesicle becomes a
prolate ellipsoid. For small eccentricity of such an ellipsoid the
free energy of the droplet is found. Then, using the known
expressions for the dipole moment of the ellipsoid, the optical
polarizabilities are calculated. This allows us to determine the
difference in refractive indices parallel and perpendicular to the
external field and thus to find the Kerr constant. In a similar
way the calculations are generalized to the droplets with finite
thickness of the surface membrane. Here we use the results due to
Van der Linden \textit{et al}.~\cite{linden}. However, our
correction of the surface free energy found in that work leads to
a significantly different expression for the Kerr constant.
Comparing our result with the experiment~\cite{linden} on a
droplet microemulsion, we estimate the bending rigidity constant
of the droplet to be about two times larger than the value
determined in the original work.

\section*{The Kerr effect on a fluid droplet}

Consider a flexible dielectric droplet taking a spherical shape in
equilibrium, immersed in another fluid. The fluid of the droplet
is assumed to be incompressible. When such a droplet of the radius
$R$ is placed in a weak electric field $\overrightarrow{E}_{0}$
directed along the axis $z$, it becomes a prolate ellipsoid with
the half-axes, to the second order of the small eccentricity
$e=\sqrt{1-b^{2}/a^{2}}$,
\begin{equation}
\label{eq1}
a=R(1+e^{2}/3),   \qquad b=c=R(1-e^{2}/6).
\end{equation}
Within the Helfrich model of interfacial
elasticity~\cite{canham,helfrich} the free energy of such an
ellipsoid (without the electrostatic energy) is~\cite{borkovec}
\begin{equation}
\label{eq2}
F=-\Delta pV+\sigma A+\int dA[\frac{\kappa}{2}(c_{1}+c_{2}-2/R_{s})^{2}+
\overline{\kappa}c_{1}c_{2}].
\end{equation}
Here $V$ is the (constant) volume of the droplet, $\Delta p$ is
the pressure inside minus outside the droplet, and $\sigma$ is the
microscopic surface tension. The integral over the surface $A$ of
the ellipsoid yields the bending energy of the droplet. It is
determined through the local curvatures $c_{1}$, $c_{2}$, and the
spontaneous curvature radius $R_{s}$. Performing the integration
over the ellipsoid, one finds
\begin{equation}
\label{eq3} F=F_{0}+\frac{8\pi}{45}e^{4}R^{2}(\alpha -
\frac{4\kappa}{RR_{s}}+\frac{6\kappa}{R^{2}}),
\end{equation}
where $F_{0}$ is for the sphere, and
$\alpha=\sigma+2\kappa/R_{s}^{2}$ is now the macroscopic surface
tension for the plane interface. The full free energy is obtained
adding the energy of the ellipsoid in the electric field $E_{0}$.
Let the dielectric constant of the ellipsoid is $\epsilon^{(i)}$,
and that of its exterior $\epsilon^{(e)}$, denoting
$\epsilon=\epsilon^{(i)}/\epsilon^{(e)}$. Then the electrostatic
energy is~\cite{landau}
\begin{equation}
\label{eq4}
F_{el}=-\frac{V\epsilon^{(e)}}{8\pi}\frac{\epsilon-1}{1+n^{(z)}(\epsilon-1)}E_{0}^{2}.
\end{equation}
The depolarization coefficient is $n^{(z)}\approx
[1-4(a-b)/5R]/3$, so that we have
\begin{equation}
\label{eq5}
F_{el}\approx-\frac{R^{3}}{2}\epsilon^{(e)}\frac{\epsilon-1}{\epsilon+2}
[1+\frac{2}{5}\frac{\epsilon-1}
{\epsilon+2}e^{2}]E_{0}^{2}.
\end{equation}
Minimalizing the full free energy with respect to the eccentricity
we find
\begin{equation}
\label{eq6}
e^{2}=\frac{9\epsilon^{(e)}}{16\pi}(\frac{\epsilon-1}{\epsilon+2})^{2}
(\alpha - \frac{4\kappa}{RR_{s}}+\frac{6\kappa}{R^{2}})^{-1}RE_{0}^{2}.
\end{equation}
To describe the Kerr birefringence, we now need the optical
polarizabilities perpendicular $\alpha_{\perp}^{opt}
=\alpha_{xx}^{opt}$ and parallel
$\alpha_{\parallel}^{opt}=\alpha_{zz}^{opt}$ to the external
field. They are obtained from the expressions for the dipole
moment of the ellipsoid with small eccentricity~\cite{landau},
\begin{equation} \label{eq7}
\alpha_{\perp}^{opt}\approx \frac{3V}{4\pi}\frac{n^{2}-1}{n^{2}+2}
[1-\frac{e^{2}}{5}\frac{n^{2}-1}{n^{2}+2}],    \qquad
\alpha_{\parallel}^{opt}\approx \frac{3V}{4\pi}\frac{n^{2}-1}
{n^{2}+2}[1+\frac{2e^{2}}{5}\frac{n^{2}-1}{n^{2}+2}],
\end{equation}
with $n$ being the relative refractive index $n=n^{(i)}/n^{(e)}$.
The difference in the refractive indices parallel and
perpendicular to the field is~\cite{jackson},
\begin{equation}
\label{eq8}
n_{\parallel}-n_{\perp}\approx 2\pi n^{(e)}\rho(\alpha_{\parallel}^{opt}-\alpha_{\perp}^{opt})
\end{equation}
($n^{(e)}\approx n_{0}$ is the refractive index of the
microemulsion when the number of droplets per unit volume, $\rho$,
is small). Using instead of $\rho$ the volume fraction of the
droplets, $\Phi$, we thus have
\begin{equation}
\label{eq9}
n_{\parallel}-n_{\perp}\approx \frac{9}{10}n_{0}(\frac{n^{2}-1}{n^{2}+2})^{2}\Phi e^{2}.
\end{equation}
The Kerr constant is defined as
\begin{equation}
\label{eq10}
K=\frac{n_{\parallel}-n_{\perp}}{E_{0}^{2}\Phi}.
\end{equation}
Substituting here the eccentricity from Eq.(6), we finally obtain
\begin{equation}
\label{eq11}
K=\frac{81}{160\pi}Rn_{0}\epsilon^{(e)}(\frac{\epsilon-1}{\epsilon+2})^{2}
(\frac{n^{2}-1}{n^{2}+2})^{2}
(\alpha - \frac{4\kappa}{RR_{s}}+\frac{6\kappa}{R^{2}})^{-1}.
\end{equation}

Note that if the entropy of mixing~\cite{ruckenstein} is taken
into account, $-kTF(\Phi)/(4\pi R^{2})$ where $F(\Phi)\approx
\ln\Phi-1$ for small $\Phi$, should be added into the last term of
Eq.(11)~\cite{lisybru}. For the case of two-phase coexistence in
microemulsions~\cite{borkovec} Eq.(11) significantly simplifies
since then $R/R_{s}=(2\kappa+\overline{\kappa})/2\kappa$ and
$\alpha=2\kappa/RR_{s}$.

\section*{The Kerr effect on a droplet covered with a shell}

The simple method used in the preceding section can be readily
generalized for the description of the Kerr effect on the droplets
covered with a vesicle membrane or a surfactant shell of nonzero
thickness. To do this we need only the expressions that generalize
Eqs.(7) for the polarizabilities of the ellipsoid, taking into
account the size of the surface shell. Such expressions have been
found in the work~\cite{linden}. Following that work we assume the
vesicle water core of radius $R_{w}$ to be characterized by the
dielectric constant $\epsilon_{w}$, and the continuous phase of
surrounding fluid by the constant $\epsilon_{o}$. For simplicity
and in order to make a comparison with the experiment, we use
$\epsilon_{w}\gg \epsilon_{o}$ which is true when $\epsilon_{o}$
stays for oil that corresponds to the experiments~\cite{linden}. A
generalization to other, more complicated cases, is
straightforward; the corresponding formulae for the
polarizabilities can be found in the work~\cite{linden}. The
surface shell can consist of two parts: a polar part which is
characterized by the constant thickness $D_{\epsilon}$, and an
apolar part of the thickness $D-D_{\epsilon}$. The polar part is
described by the dielectric constant $\epsilon_{\beta}$,
characterizing the orthogonal (to the surface) components of the
dielectric constant, and by $\epsilon_{\gamma}$ for the parallel
components. Both $\epsilon_{\beta}$ and $\epsilon_{\gamma}$ are
large compared to $\epsilon_{o}$. The apolar part of the layer has
the dielectric constant approximately the same as for the oil.
Then the parallel component of the polarizability tensor of such
an ellipsoid is as follows~\cite{linden}:$$\alpha_{\parallel}=
R_{w}^{3}\frac{\epsilon_{w}-\epsilon_{o}}{\epsilon_{w}+2\epsilon_{o}}[1+3ae^{2}
\frac{\epsilon_{w}-\epsilon_{o}}{\epsilon_{w}+2\epsilon_{o}}]
+R_{w}^{2}D_{\epsilon}\frac{3\epsilon_{o}}{(\epsilon_{w}+2\epsilon_{o})^{2}}$$
\begin{equation}
\label{eq12}
\times\{{\epsilon_{w}^{2}(\frac{1}{\epsilon_{o}}-\frac{1}{\epsilon_{\beta}})[1+2ae^{2}
\frac{\epsilon_{w}-7\epsilon_{o}}{\epsilon_{w}+2\epsilon_{o}}]+2(\epsilon_{\gamma}-
\epsilon_{o})[1+2ae^{2}\frac{4\epsilon_{w}-\epsilon_{o}}{\epsilon_{w}+2\epsilon_{o}}]}\},
\end{equation}
where $a=2/15$. For $\alpha_{\perp}$ the same expression is valid
but with $a=-1/15$. In the calculation of the eccentricity $e$ we
use the above mentioned inequalities for the dielectric constants
that gives
\begin{equation}
\label{eq13} \alpha_{\parallel}\approx
R_{w}^{3}(1+\frac{2}{5}e^{2})+3R_{w}^{2}D_{\epsilon}
[1+\frac{4}{15}e^{2}+2\frac{\epsilon_{o}\epsilon_{\gamma}}{\epsilon_{w}^{2}}
(1+\frac{16}{15}e^{2})].
\end{equation}
Using this expression we find the electrostatic part of the free
energy of the ellipsoid, which is now instead of Eq.(4)
\begin{equation}
\label{eq14}
F_{el}=-\frac{1}{2}\alpha_{\parallel}\epsilon_{o}E_{0}^{2}.
\end{equation}
Minimalizing the full free energy $F+F_{el}$, with $F$ from
Eq.(3), the eccentricity is
\begin{equation}
\label{eq15}
e^{2}\approx\frac{9\epsilon_{o}R_{w}E_{0}^{2}}{16\pi}(\alpha - \frac{4\kappa}
{R_{w}R_{s}}+\frac{6\kappa}{R_{w}^{2}})^{-1}
(1+2\frac{D_{\epsilon}}{R_{w}}).
\end{equation}
From Eq.(8), rewriting the polarizabilities $\alpha_{\parallel}$
and $\alpha_{\perp}$ from Eq.(12) for the optical case simply
changing the static dielectric constants by the squares of the
refractive indices~\cite{linden}, we finally obtain$$
\frac{n_{\parallel}-n_{\perp}}{E_{0}^{2}}\approx
\frac{27}{40}R_{w}^{4}\rho n_{o}\epsilon_{o} (\alpha -
\frac{4\kappa}{R_{w}R_{s}}+\frac{6\kappa}{R_{w}^{2}})^{-1}(1+2\frac{D_{\epsilon}}{R_{w}})
\{(\frac{n_{w}^{2}-n_{0}^{2}}{n_{w}^{2}+2n_{o}^{2}})^{2}$$
\begin{equation}
\label{eq16}
+\frac{2Dn_{o}^{2}}{R_{w}(n_{w}^{2}+2n_{o}^{2})^{3}}[n_{w}^{4}(n_{o}^{-2}-n_{\beta}^{-2})(n_{w}^{2}-7n_{o}^{2})
+2(n_{\gamma}^{2}-n_{o}^{2})(4n_{w}^{2}-n_{o}^{2})]\}.
\end{equation}
The difference between this result and the result by Van der
Linden \textit{et al.}~\cite{linden} is significant. This is
because of the difference in the surface energy of the deformed
droplet in the electric field: they have in the first bracket in
Eq.(16) only the term $6\kappa/R_{w}^{2}$. The rest terms are,
however, not negligible if compared with this one: for a detailed
discussion we can refer, e.g., to the work~\cite{ruckenstein}.
Neglecting the surface energy associated with the surface tension
is justified only for an absolutely free vesicle membrane with
identical fluids inside and outside it, but not in other cases. As
well, in general one cannot assume $R_{w}/R_{s}\ll1$ and drop out
the corresponding terms as it was done in~\cite{linden}. So, for a
microemulsion droplet, the two radii can be comparable, e.g. in
the case of the so-called two-phase coexistence we have
$R_{w}/R_{s}\approx (2\kappa+\overline{\kappa})/2\kappa$.\\
In the paper~\cite{linden} a detailed comparison between the
theory and the Kerr effect experiment was done from which the
value $\kappa\approx 0.46 kT$ has been extracted. Taking into
account the different units used and the above discussed
improvement of the theory, it is seen that this value of the
rigidity constant is essentially underestimated. Really, let us
express~\cite{lisybru} in Eq.(16)$$\frac{\alpha R_{w}^{2}}{6} -
\frac{2\kappa R_{w}}{3R_{s}}+\kappa\approx
(\frac{\kappa}{kT}-\frac{1}{48\pi\varepsilon})kT,$$ where
$\varepsilon$ is the polydispersity of the droplets in radii.
Exactly this expression should be used in the analysis of the
experimental data~\cite{linden} that yielded the value
$\kappa\approx 0.46 kT$. Taking into account the polydispersity in
the sample, one can see that the lower the polydispersity is, the
higher value of $\kappa$ would be determined from the experiment.
For example, for a typical polydispersity index~\cite{erratum}
$\sqrt{\varepsilon}=0.12$ one obtains $\kappa\approx 0.92 kT$: a
value two times larger than that found in Ref.~\cite{linden} for
the water-AOT-isooctane droplet microemulsion and very close to
that determined from the Kerr effect measurements by Borkovec and
Eicke~\cite{borkeicke,erratum}.

\section*{Conclusion}

The Kerr effect measurements on vesicle-like droplets could in
principle serve as an alternative (to such techniques like the
static and dynamic neutron or light scattering) probe of the
properties of the droplet surface membranes or films. In the
present work the Kerr effect on dilute suspension of such
(noninteracting) droplets has been described. We evaluated the
Kerr constant of a fluid droplet immersed in another fluid
assuming that the droplet in a constant weak electric field takes
a shape of the prolate ellipsoid. This assumption is substantiated
by the works~\cite{bork,lisy} where the flexible dielectric sphere
in the electric field is considered coming from the Laplace
equation for the electric potential with the appropriate boundary
conditions. It was shown that in the lowest approximation the
sphere undergoes ellipsoidal fluctuations and after the
statistical averaging in the presence of the electric field the
result for the Kerr constant is the same as for a prolate
axisymmetric ellipsoid with small eccentricity. Following the
work~\cite{linden} we obtained a very different result for the
Kerr constant that is due to the different expression found for
the surface elastic energy of the droplet. As a result, the Kerr
constant extracted from the experimental data~\cite{linden}
differs significantly, about two times, from the value determined
in the original work. However, also this value should be
considered with some doubts. First, the experimental error in
obtaining the Kerr constant by extrapolation of the data to zero
concentration of the droplets is large so that the estimation is
not very reliable. The signal detected in the Kerr-effect
measurements is very sensitive to the droplet radius, so it should
be measured in the experiments with high precision together with
the polydispersity of the droplets in radii, that was not
determined in the described experiments at all. Finally, the
recent investigation~\cite{edwards} suggests that a relevant
theory of the electro-optical measurements on solutions of
vesicle-like droplets should incorporate many-particle
correlations even in the case of small concentrations of the
droplets. It appears that long-range anisotropic density
correlations resulting from dipolar interactions have to be taken
into account in a generalization of the simple single-body theory
presented here.
\\

\textbf{Acknowledgment}. This work is a part of the PhD. thesis by
M\'{a}ria Richterov\'{a}. It was supported by the grant VEGA No.
1/7401/00, Slovak Republic.

\end{document}